\documentclass[11pt]{article}

\usepackage{fullpage}
\usepackage{latexsym}
\usepackage{amsfonts}
\usepackage{amssymb}
\usepackage{amsmath}
\usepackage{color}
\usepackage{graphicx, graphics}
\usepackage{hyperref}
\usepackage{pstricks, pst-plot, epsfig}

\newcommand{\beq}{\begin{equation}}
\newcommand{\eeq}{\end{equation}}
\newcommand{\ba}{\begin{array}}
\newcommand{\ea}{\end{array}}
\newcommand{\bea}{\begin{eqnarray*}}
\newcommand{\eea}{\end{eqnarray*}}
\newcommand{\bc}{\begin{center}}
\newcommand{\ec}{\end{center}}
\newcommand{\bt}{\begin{table}}
\newcommand{\et}{\end{table}}

\newcommand{\sgn}{\mbox{sgn}}

\newcommand{\no}{\noindent}

\newcommand{\beqno}{\begin{displaymath}}
\newcommand{\eeqno}{\end{displaymath}}

\newcommand{\been}{\begin{enumerate}}
\newcommand{\een}{\end{enumerate}}

\newcommand{\ud}{\,\mathrm{d}}
\renewcommand\Re{\operatorname{Re}}
\renewcommand\Im{\operatorname{Im}}

\newcommand{\CC}{\mathbb{C}}

\newcommand{\edits}[1]{\textcolor{red}{#1}}
\renewcommand{\edits}[1]{\textcolor{black}{#1}}

\newtheorem{prop}{Proposition}

\begin{document}

\title{Interface Problems for Dispersive equations}
\author{
Natalie E Sheils~~~~Bernard Deconinck\\
~\\
Department of Applied Mathematics\\
University of Washington\\
Seattle, WA 98195-2420\\
nsheils{@}amath.washington.edu~~bernard{@}amath.washington.edu}

\maketitle

\begin{abstract}
The interface problem for the linear Schr\"odinger equation in one-dimensional piecewise homogeneous domains is examined by providing an explicit solution in each domain. The location of the interfaces is known and the continuity of the wave function and a jump in their derivative at the interface are the only conditions imposed. The problem of two semi-infinite domains and that of two finite-sized domains are examined in detail.  The problem and the method considered here extend that of an earlier paper by Deconinck, Pelloni and Sheils (2014).  The dispersive nature of the problem presents additional difficulties that are addressed here. 
\end{abstract}

\section{Introduction}

Interface problems for partial differential equations (PDEs) are initial boundary value problems for which the solution of an equation in one domain prescribes boundary conditions for the equations in adjacent domains. In applications, precise interface conditions follow from conservations laws. Few interface problems allow for an explicit closed-form solution using classical solution methods. Using the Fokas method ~\cite{DeconinckTrogdonVasan, FokasBook, FokasPelloni4} such solutions may be constructed for both dissipative and dispersive linear interface problems. 

In two recent papers~\cite{Asvestas, DeconinckPelloniSheils} this was done for the classical problem of the heat equation.  In~\cite{DeconinckPelloniSheils} the main application considered is that of heat flow in composite walls or rods while in~\cite{Asvestas} the heat equation is viewed as a simplified reaction-diffusion equation describing the spreading of tumors in the brain.  Problems in both finite and infinite domains were investigated in~\cite{DeconinckPelloniSheils} and the method was compared with classical solution approaches if such exist~\cite{CarslawJaeger, HahnO}.  The same is done here for the linear Schr\"odinger (LS) equation with an interface. We restrict to the case of a continuous wave function with a jump in the derivative across the interface.  Although the problem and the method considered here are similar to the one presented in~\cite{Asvestas} and~\cite{DeconinckPelloniSheils}, the dispersive nature of the problem makes it more difficult to solve both classically and using the method of Fokas. 

The linear Schr\"odinger equation is arguably the simplest dispersive equation, having the dispersion relation $\omega(k)=k^2$. It arises in its own right in quantum mechanics~\cite{Schrodinger}, and as the linearization of various nonlinear equations, most notably the nonlinear Schr\"odinger (NLS) equations $iq_t(x,t)=-q_{xx}(x,t) \pm  |q(x,t)|^2 q(x,t)$. As such, it arises in a large variety of application areas, whenever the modulation of nonlinear wave trains is considered. Indeed, it has
been derived in such diverse fields as waves in deep water~\cite{Zak}, plasma physics~\cite{Zak2}, nonlinear fiber optics~\cite{HasegawaTappert1, HasegawaTappert2}, magneto-static spin waves~\cite{ZvezdinPopkov}, and many other settings.

The LS equation describes the behavior of solutions of the NLS equation in the small amplitude limit and understanding it dynamics is fundamental in understanding the dynamics of the more complicated nonlinear problem.

Recently Cascaval and Hunter~\cite{CascavalHunter} have considered the time-dependent LS on simple networks.  Their solution formulas are not explicit, as they contain implicit integral equations for the interface conditions. Their analysis is easily extended to more than two domains and also considers the nonlinear Schr\"odinger (NLS) equation. \edits{Some work has been done using the Fokas Method for moving boundary value problems in the case when the movement of the boundary is prescribed~\cite{FokasPelloni5}.  In some cases, the solution of such problems requires the use of the ``d-bar method" which reduces the problem to a linear integral equation.}

The LS equation in two semi-infinite domains with an interface is considered in Section~\ref{sec:ii}. The method is adapted to the problem of two finite domains in Section~\ref{sec:ff}.  The solution formulae given are easily computed numerically using techniques presented in~\cite{Levin, TrogdonThesis}.  Throughout, our emphasis is on non-steady state solutions.  The solutions presented here using the Fokas Method are explicit and depend only on known quantities.  Although we present solution formulas only for the case of two domains (both finite or both infinite) it is straightforward to generalize this method to $n$ domains.  This is done explicitly for multiple domains for the heat equation in~\cite{DeconinckPelloniSheils} (three domains, both finite and infinite) and in~\cite{Asvestas} ($n$ finite domains) and the process here would be similar.

\section{Two semi-infinite domains}\label{sec:ii}

We wish to find $q^L(x,t)$ and $q^R(x,t)$ satisfying
\beq\label{2ieqns}
\begin{aligned}
i q^L_t(x,t)=&\sigma_Lq^L_{xx}(x,t),~~~&-\infty<&x<0,~&t>0,\\
i q^R_t(x,t)=&\sigma_Rq^R_{xx}(x,t),~~~ &0<&x<\infty,~&t>0,
\end{aligned}
\eeq
subject to the asymptotic conditions 
\beq\label{2i_bc}
\begin{aligned}
\lim_{x\to-\infty}q^L(x,t)=&\gamma_L,\\
\lim_{x\to\infty}q^R(x,t)=&\gamma_R,
\end{aligned}
\eeq
the initial conditions
\beq
\begin{aligned}
q^L(x,0)=&q^L_0(x),~~~&-\infty&<x<0,\\
q^R(x,0)=&q^R_0(x),~~~&0&<x<\infty,
\end{aligned}
\eeq
and the interface conditions 
\beq\label{2i_ifc}
\begin{aligned}
q^L(0,t)=&q^R(0,t),~~~&t>0,\\
\beta_L q^L_x(0,t)=&\beta_R q^R_x(0,t),~~~&t>0,
\end{aligned}
\eeq
where $\gamma_L,\gamma_R,\sigma_L,\sigma_R, \beta_L$ and $\beta_R$ are $t$-independent nonzero constants. The sub- and super-indices $L$ and $R$ denote the left and right domain, respectively.  In what follows we assume that $\sigma_L$ and $\sigma_R$ are both positive for convenience.  

First, we shift the problem so that the asymptotic conditions are identically zero.  We define $v^L(x,t)=q^L(x,t)-\gamma_L$ and $v^R(x,t)=q^R(x,t)-\gamma_R$ that satisfy
\begin{subequations}
\begin{align}
iv^L_t(x,t)=&\sigma_Lv^L_{xx}(x,t),&-\infty<&x<0,&&t\geq0,\label{LS_L}\\
iv^R_t(x,t)=&\sigma_Rv^R_{xx}(x,t),&0<&x<\infty,&&t\geq0,\label{LS_R} \\
\lim_{x\to-\infty}v^L(x,t)=&0,&&t\geq0,\\
\lim_{x\to\infty}v^R(x,t)=&0,&&t\geq 0,\\
v^L(x,0)=&v^L_0(x),&-\infty<&x<0,\\
v^R(x,0)=&v^R_0(x),&0<&x<\infty,\\
v^L(0,t)+\gamma^L=&v^R(0,t)+\gamma^R,&&t\geq 0,\\
\beta_Lv^L_x(0,t)=&\beta_Rv^R_x(0,t),&&t\geq 0.
\end{align}
\end{subequations}

We follow the standard steps in the application of the Fokas Method~\cite{DeconinckTrogdonVasan, FokasBook, FokasPelloni4}, beginning with the so-called ``local relations''~\cite{DeconinckTrogdonVasan}

\begin{subequations}
\begin{align}
\label{local2iL}
\left(e^{-ikx+\edits{\lambda_L}t}v^L(x,t)\right)_t=&\left(\sigma_Le^{-ikx+\edits{\lambda_L}t}(kv^L(x,t)-iv^L_x(x,t))\right)_x,~~~&-\infty<&x<0,\\
\label{local2iR}
\left(e^{-ikx+\edits{\lambda_R}t}v^R(x,t)\right)_t=&\left(\sigma_Re^{-ikx+\edits{\lambda_R}t}(kv^R(x,t)-iv^R_x(x,t))\right)_x,~~~&0<&x<\infty.
\end{align}
\end{subequations}

\begin{figure}
\begin{center}
\def\svgwidth{4in}
   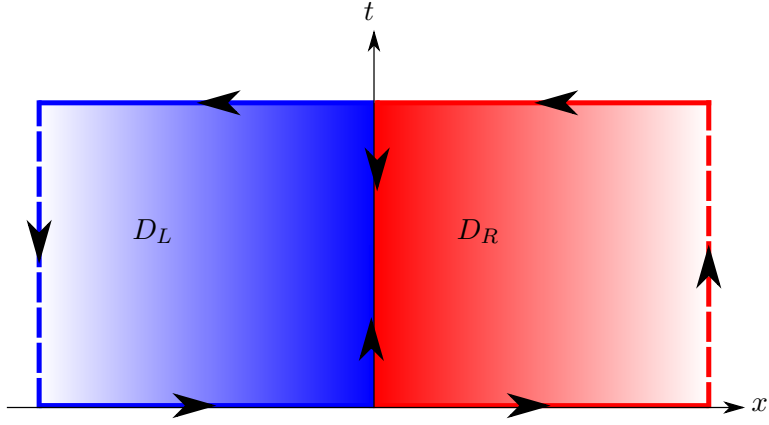 
   \caption{Domains for the application of Green's Theorem for $v^L(x,t)$ and $v^R(x,t)$.   \label{fig:LHP_RHP}}
  \end{center}
\end{figure}

These are parameter relations obtained by rewriting \eqref{LS_L} and \eqref{LS_R}. This also tell us $\edits{\lambda_L}=-i\sigma_Lk^2$ and $\edits{\lambda_R}=-i\sigma_Rk^2$.  \edits{The functions $\lambda_L(k)$ and $\lambda_R(k)$ are related to the dispersion relations of the equations: $\lambda_L=-i \omega_L$ and $\lambda_R=-i \omega_R$.  The use of these functions instead of the dispersion relations proper is common when using the Fokas method and we continue this use.}  Applying Green's Theorem~\cite{AblowitzFokas} in the strip $(-\infty,0)\times (0,t)$, see Figure~\ref{fig:LHP_RHP}, we find the global relations

\begin{subequations}\label{global2i}
\begin{align}
\label{global2iL}
0=&\int_{-\infty}^0 e^{-ikx} v_0^L(x)\ud x-\int_{-\infty}^0 e^{-ikx+\edits{\lambda_L} t}v^L(x,t)\ud x+\int_0^t \sigma_Le^{\edits{\lambda_L} s}\left( k v^L(0,s)-iv^L_x(0,s)  \right)\ud s,\\
\label{global2iR}
0=&\int_0^\infty e^{-ikx} v_0^R(x)\ud x-\int_0^\infty e^{-ikx+\edits{\lambda_R} t}v^R(x,t)\ud x-\int_0^t \sigma_Re^{\edits{\lambda_R} s}\left( k v^R(0,s)-iv^R_x(0,s)  \right)\ud s.
\end{align}
\end{subequations}

Let $\CC^+=\{z\in\CC:\Im(z)\geq0\}$.  Similarly, let $\CC^-=\{z\in\CC:\Im(z)\leq0\}$.  The Fourier integrals in~\eqref{global2i} require $k\in\CC^+$ in~\eqref{global2iL} and $k\in\CC^-$ in~\eqref{global2iR}.  For $k\in\CC$, we define the following transforms:

\begin{align*}
g_{0}({\omega},t)=&\int_{0}^te^{\omega s}v^L(0,s)\ud s=\int_{0}^te^{\omega s}(v^R(0,s)+\gamma^R-\gamma^L)\ud s\\
=&\frac{(\gamma^R-\gamma^L)(e^{\omega t}-1)}{\omega}+\int_{0}^te^{\omega s}v^R(0,s)\ud s,\\
g_{1}({\omega},t)=&\int_{0}^te^{{\omega} s}v^L_x(0,s)\ud s=\frac{\beta_R}{\beta_L} \int_{0}^te^{{\omega} s}v^R_x(0,s)\ud s,\\
\hat{v}^L(k,t)=&\int_{-\infty}^0e^{-ikx}v^L(x,t)\ud x, &\hat{v}^L_0(k)=&\int_{-\infty}^0e^{-ikx}v^L_0(x)\ud x,\\
\hat{v}^R(k,t)=&\int_{0}^\infty e^{-ikx}v^R(x,t)\ud x, &\hat{v}^R_0(k)=&\int_{0}^\infty e^{-ikx}v^R_0(x)\ud x.
\end{align*}

\no Using these definitions, the global relations~\eqref{global2i} are rewritten as
\begin{subequations}\label{tglobal2i}
\begin{align}
\label{tglobal2iL}
0=&\hat{v}_0^L(k)-e^{\edits{\lambda_L} t}\hat{v}^L(k,t)+k\sigma_Lg_0(\edits{\lambda_L},t)-i\sigma_Lg_1(\edits{\lambda_L},t),\\
\label{tglobal2iR}
0=&\hat{v}_0^R(k)-e^{\edits{\lambda_R}t}\hat{v}^R(k,t)-k\sigma_Rg_0(\edits{\lambda_R},t) +\frac{i(\gamma_R-\gamma_L)}{k}(e^{\edits{\lambda_R}t}-1)+\frac{i\sigma_R\beta_L}{\beta_R}g_1(\edits{\lambda_R},t),
\end{align}
\end{subequations}
where $k\in\CC^+$ for~\eqref{tglobal2iL} and $k\in\CC^-$ for~\eqref{tglobal2iR}. Since \edits{the functions $\lambda_R$ and $\lambda_L$} are invariant under $k\to-k$ we can supplement~\eqref{tglobal2i} with their evaluation at $-k$, namely
\begin{subequations}\label{tglobal2i minus}
\begin{align}
\label{tglobal2iL minus}
0=&\hat{v}_0^L(-k)-e^{\edits{\lambda_L} t}\hat{v}^L(-k,t)-k\sigma_Lg_0(\edits{\lambda_L},t)-i\sigma_Lg_1(\edits{\lambda_L},t),\\
\label{tglobal2iR minus}
0=&\hat{v}_0^R(-k)-e^{\edits{\lambda_R}t}\hat{v}^R(-k,t)+k\sigma_Rg_0(\edits{\lambda_R},t) -\frac{i(\gamma_R-\gamma_L)}{k}(e^{\edits{\lambda_R}t}-1)+\frac{i\sigma_R\beta_L}{\beta_R}g_1(\edits{\lambda_R},t),
\end{align}
\end{subequations}
where $k\in\CC^-$ for~\eqref{tglobal2iL minus} and $k\in\CC^+$ for~\eqref{tglobal2iR minus}.

Inverting the Fourier transform in~\eqref{tglobal2iL} we have
$$
v^L(x,t)=\frac{1}{2\pi}\int_{-\infty}^\infty e^{ikx-\edits{\lambda_L}t}\hat{v}_0^L(k)\ud k+\frac{1}{2\pi}\int_{-\infty}^\infty e^{ikx-\edits{\lambda_L}t}\sigma_L(kg_0(\edits{\lambda_L},t)-ig_1(\edits{\lambda_L},t))\ud k,
$$

\begin{figure}
\begin{center}
\def\svgwidth{4in}
   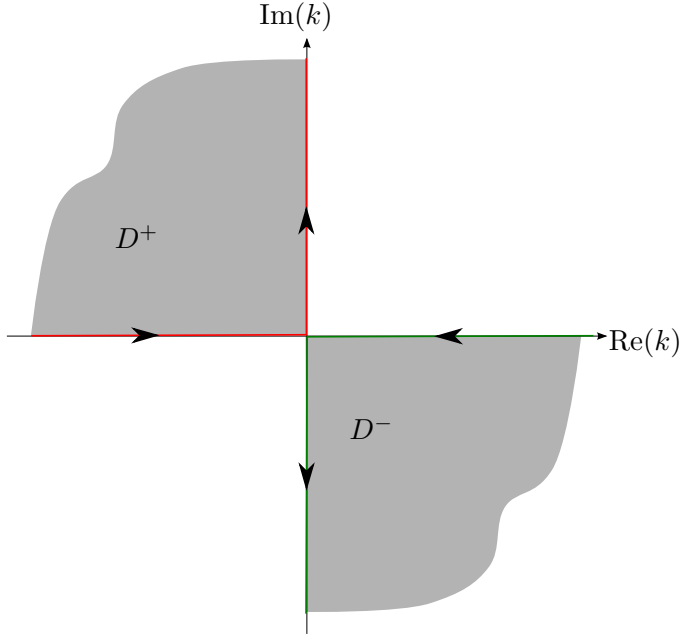 
   \caption{The Domains $D^+$ and $D^-$ for the linear Schr\"odinger equation.   \label{fig:LS}}
   \end{center}
\end{figure}

\no for $-\infty<x<0$ and $t>0$. Let $D=\{k\in\CC:\Re({-ik^2})<0\}=D^+\cup D^-$.  The region $D$ is shown in Figure~\ref{fig:LS}.  The integrand of the second integral is entire and decays as $k\to\infty$ for $k\in \CC^-\setminus D^-$.  Using the analyticity of the integrand and applying Jordan's Lemma~\cite{AblowitzFokas} we can replace the contour of integration of the second integral by $-\int_{\partial D^-}$:

\beq\label{soln2iL}
v^L(x,t)=\frac{1}{2\pi}\int_{-\infty}^\infty e^{ikx-\edits{\lambda_L}t}\hat{v}_0^L(k)\ud k-\frac{1}{2\pi}\int_{\partial D^-} e^{ikx-\edits{\lambda_L}t}\sigma_L(kg_0(\edits{\lambda_L},t)-ig_1(\edits{\lambda_L},t))\ud k.
\eeq

Similarly, inverting the Fourier transform in~\eqref{tglobal2iR} we have
\begin{equation*}
\begin{split}
v^R(x,t)
=&\frac{(\gamma_R-\gamma_L)}{2}\phi(\sigma_R,x,t)+\frac{1}{2\pi}\int_{-\infty}^\infty e^{ikx-\edits{\lambda_R} t}\hat{v}_0^R(k)\ud k\\
&+\frac{1}{2\pi}\int_{-\infty}^\infty e^{ikx-\edits{\lambda_R} t}\sigma_R\left(-kg_0(\edits{\lambda_R},t)+\frac{i\beta_L}{\beta_R}g_1(\edits{\lambda_R},t)\right)\ud k,
\end{split}
\end{equation*}

\no for $0<x<\infty$, $t>0$, and $$\phi(\sigma,x,t)=\left\{ \ba{lcr} 0,&&t=0,\\ -\sgn(x)+\frac{1}{\sqrt{\pi\sigma t}}e^{i\pi/4}\int_0^x e^{-iy^2/(4\sigma t)}\ud y,&&t>0.\ea\right.$$  The integrand of the third integral is entire and decays as $k\to\infty$ for $k\in \CC^+\setminus D^+$.  Using the analyticity of the integrand and applying Jordan's Lemma~\cite{AblowitzFokas} we can replace the contour of integration of this integral by $\int_{\partial D^+}$:
\beq\label{soln2iR}
\begin{split}
v^R(x,t)=&\frac{\gamma_R-\gamma_L}{2}\phi(\sigma_R,x,t)+\frac{1}{2\pi}\int_{-\infty}^\infty e^{ikx-\edits{\lambda_R} t}\hat{v}_0^R(k)\ud k\\
&+\frac{1}{2\pi}\int_{\partial D^+}e^{ikx-\edits{\lambda_R} t}\sigma_R\left(-kg_0(\edits{\lambda_R},t)+\frac{i\beta_L}{\beta_R}g_1(\edits{\lambda_R},t)\right)\ud k.
\end{split}
\eeq

The expressions \eqref{soln2iL} and \eqref{soln2iR} for $v^L(x,t)$ and $v^R(x,t)$ depend on the unknown functions $g_0$ and $g_1$, evaluated at different arguments. These functions need to be expressed in terms of known quantities.  To obtain a system of two equations for the two unknown functions we use the four global relations.  We use \eqref{tglobal2iR} and \eqref{tglobal2iL minus} for $g_0(\edits{\lambda_L},t)$, and $g_1(\edits{\lambda_L},t)$. This requires use of all the symmetries of the set of \edits{$\{\lambda_L(k), \lambda_R(k)\}$.}  Namely, the transformation $k\to \sqrt{\sigma_L/\sigma_R}k$ in \eqref{tglobal2iR}.   Substituting these into~\eqref{soln2iL} we have

\begingroup
\setlength{\thinmuskip}{0mu}
\setlength{\medmuskip}{0mu}
\setlength{\thickmuskip}{0mu}
\beq\label{soln2iLfull}
\begin{split}
v^L(x,t)
=&\frac{\beta_R\sigma_L(\gamma_R-\gamma_L)}{\beta_R\sigma_L+\beta_L\sqrt{\sigma_L\sigma_R}}\phi(\sigma_L,x,t)+\frac{1}{2\pi}\int_{-\infty}^\infty e^{ikx-\edits{\lambda_L}t}\hat{v}_0^L(k)\ud k\\
&+\frac{\beta_R\sigma_L-\beta_L\sqrt{\sigma_L\sigma_R}}{2\pi(\beta_R \sigma_L+\beta_L\sqrt{\sigma_L\sigma_R})}\int_{\partial D^-} e^{ikx-\edits{\lambda_L}t}\hat{v}_0^L(-k)\ud k\\
&-\frac{\beta_R\sigma_L}{\pi(\sigma_R\beta_L+\beta_R\sqrt{\sigma_L\sigma_R})}\int_{\partial D^-} e^{ikx-\edits{\lambda_L}t}\hat{v}_0^R\left(k\sqrt{\frac{\sigma_L}{\sigma_R}}\right)\ud k\\
&-\frac{\beta_R\sigma_L-\beta_L\sqrt{\sigma_L\sigma_R}}{2\pi(\beta_R \sigma_L+\beta_L\sqrt{\sigma_L\sigma_R})}\int_{\partial D^-} e^{ikx}\hat{v}^L(-k,t)\ud k\\
&+\frac{\beta_R\sigma_L}{\pi(\sigma_R\beta_L+\beta_R\sqrt{\sigma_L\sigma_R})}\int_{\partial D^-} e^{ikx}\hat{v}^R\left(k\sqrt{\frac{\sigma_L}{\sigma_R}},t\right)\ud k,
\end{split}
\eeq
\endgroup

\begin{figure}[tb]
   \centering
\def\svgwidth{4in}
   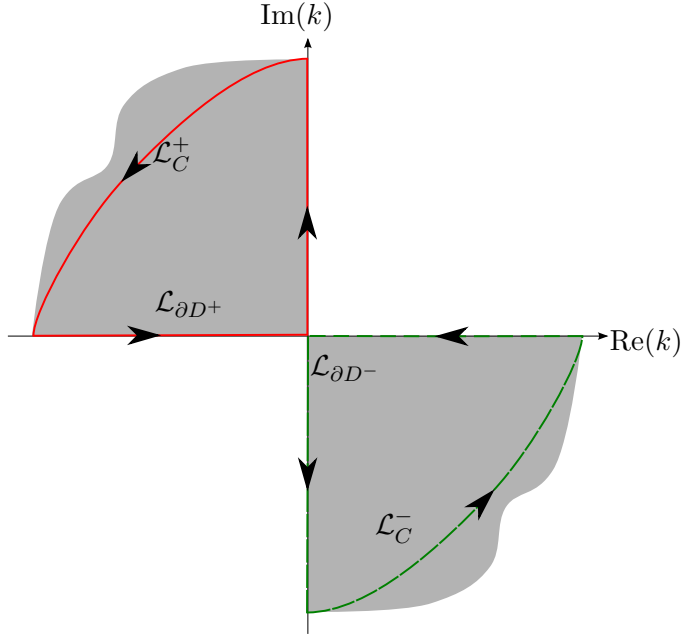 
   \caption{The contour $\mathcal{L}^-$ is shown in green as a dashed line.  An application of Cauchy's Integral Theorem ~\cite{AblowitzFokas} using this contour allows elimination of the contribution of $\hat{v}^L(-k,t)$ from the integral~\eqref{soln2iLfull}.  Similarly, the contour $\mathcal{L^+}$ is shown in red and application of Cauchy's Integral Theorem using this contour allows elimination of the contribution of $\hat{v}^R(-k,t)$ from~\eqref{soln2iRfull}.   \label{fig:LS_close}}
\end{figure}

\no for $-\infty<x<0$, $t>0$.  The first four terms depend only on known functions.  The integrand of the second-to-last term is analytic for all $k\in \CC^-$.  Further, $\hat{v}^L(-k,t)$ decays for $k\to\infty$ for $k\in\CC^-$.  Thus, by Jordan's Lemma, the integral of $\exp(ikx)\hat{v}^L(-k,t)$ along a closed, bounded curve in $\CC^-$ vanishes. In particular we consider the closed curve $\mathcal{L}^-=\mathcal{L}_{\partial D^-}\cup\mathcal{L}^-_C$ where $\mathcal{L}_{\partial D^-}=\partial D^- \cap \{k: |k|<C\}$ and $\mathcal{L}^-_C=\{k\in D^-: |k|=C\}$, see Figure~\ref{fig:LS_close}.

Since the integral along $\mathcal{L}_C^-$ vanishes as $C\to\infty$, the fourth integral on the right-hand side of \eqref{soln2iLfull} must vanish since the contour $\mathcal{L}_{\partial D^-}$  becomes $\partial D^-$ as $C\to\infty$.  The uniform decay of $\hat{v}^L(-k,t)$ for large $k$ is exactly the condition required for the integral to vanish, using Jordan's Lemma.  For the final integral in~\eqref{soln2iLfull} we use that $\hat{v}^R(k \sqrt{\sigma_L/\sigma_R},t)$ is analytic and bounded for $k\in\CC^-$.  Using the same argument as above, the fifth integral in~\eqref{soln2iLfull} vanishes and we have an explicit representation for $v^L(x,t)$ in terms of initial conditions:

\beq\label{FULLsoln2iL}
\begin{split}
v^L(x,t)=&\frac{\beta_R\sigma_L(\gamma_R-\gamma_L)}{\beta_R\sigma_L+\beta_L\sqrt{\sigma_L\sigma_R}}\phi(\sigma_L,x,t)+\frac{1}{2\pi}\int_{-\infty}^\infty e^{ikx-\edits{\lambda_L}t}\hat{v}_0^L(k)\ud k\\
&+\frac{\beta_R\sigma_L-\beta_L\sqrt{\sigma_L\sigma_R}}{2\pi(\beta_R \sigma_L+\beta_L\sqrt{\sigma_L\sigma_R})}\int_{\partial D^-} e^{ikx-\edits{\lambda_L}t}\hat{v}_0^L(-k)\ud k\\
&-\frac{\beta_R\sigma_L}{\pi(\sigma_R\beta_L+\beta_R\sqrt{\sigma_L\sigma_R})}\int_{\partial D^-} e^{ikx-\edits{\lambda_L}t}\hat{v}_0^R\left(k\sqrt{\frac{\sigma_L}{\sigma_R}}\right)\ud k,
\end{split}
\eeq

To find an explicit expression for $v^R(x,t)$ we need to evaluate $g_0$ and $g_1$ at different arguments, also ensuring that the expressions are valid for $k\in \CC^+\setminus D^+$.  Substituting these into equation~\eqref{soln2iR}, we obtain

\beq\label{soln2iRfull}
\begin{split}
v^R(x,t)=&\frac{\beta_L\sqrt{\sigma_L\sigma_R}(\gamma_R-\gamma_L)}{\beta_R\sigma_L+\beta_L\sqrt{\sigma_R\sigma_L}}\phi(\sigma_R,x,t)+\frac{1}{2\pi}\int_{-\infty}^\infty e^{ikx-\edits{\lambda_R} t}\hat{v}_0^R(k)\ud k\\
&+\frac{\beta_L\sigma_R}{\pi\left(\beta_R\sigma_L+\beta_L\sqrt{\sigma_R\sigma_L}\right)}\int_{\partial D^+}e^{ikx-\edits{\lambda_R}t}\hat{v}_0^L\left(k\sqrt{\frac{\sigma_R}{\sigma_L}}\right)\ud k\\
&+\frac{\beta_R\sigma_L-\beta_L\sqrt{\sigma_L\sigma_R}}{2\pi\left(\beta_R\sigma_L+\beta_L\sqrt{\sigma_R\sigma_L}\right)}\int_{\partial D^+}e^{ikx-\edits{\lambda_R}t}\hat{v}_0^R(-k)\ud k\\
&-\frac{\beta_L\sigma_R}{\pi\left(\beta_R\sigma_L+\beta_L\sqrt{\sigma_R\sigma_L}\right)}\int_{\partial D^+}e^{ikx}\hat{v}^L\left(k\sqrt{\frac{\sigma_R}{\sigma_L}},t\right)\ud k\\
&-\frac{\beta_R\sigma_L-\beta_L\sqrt{\sigma_L\sigma_R}}{2\pi\left(\beta_R\sigma_L+\beta_L\sqrt{\sigma_R\sigma_L}\right)}\int_{\partial D^+}e^{ikx}\hat{v}^R(-k,t)\ud k\\
\end{split}
\eeq
\no for $0<x<\infty$, $t>0$.  As before, the first four integrals are known.  To compute the fifth and sixth integrals we proceed as we did for $v^L(x,t)$ and eliminate integrals that decay in the regions over which we are integrating.  The final solution is

\beq\label{FULLsoln2iR}
\begin{split}
v^R(x,t)=&\frac{\beta_L\sqrt{\sigma_L\sigma_R}(\gamma_R-\gamma_L)}{\beta_R\sigma_L+\beta_L\sqrt{\sigma_R\sigma_L}}\phi(\sigma_R,x,t)+\frac{1}{2\pi}\int_{-\infty}^\infty e^{ikx-\edits{\lambda_R} t}\hat{v}_0^R(k)\ud k\\
&+\frac{\beta_L\sigma_R}{\pi\left(\beta_R\sigma_L+\beta_L\sqrt{\sigma_R\sigma_L}\right)}\int_{\partial D^+}e^{ikx-\edits{\lambda_R}t}\hat{v}_0^L\left(k\sqrt{\frac{\sigma_R}{\sigma_L}}\right)\ud k\\
&+\frac{\beta_R\sigma_L-\beta_L\sqrt{\sigma_L\sigma_R}}{2\pi\left(\beta_R\sigma_L+\beta_L\sqrt{\sigma_R\sigma_L}\right)}\int_{\partial D^+}e^{ikx-\edits{\lambda_R}t}\hat{v}_0^R(-k)\ud k.
\end{split}
\eeq

Returning to the original variables we have the following proposition which determines $q^R$ and $q^L$ fully explicitly in terms of the given initial conditions and the prescribed boundary conditions as $|x|\to\infty$.

\begin{prop}\label{2iprop}
The solution of the linear Schr\"odinger problem \eqref{2ieqns}-\eqref{2i_ifc} is given by
\beq\label{2iqsolnsL}
\begin{split}
q^L(x,t)=&\gamma^L+\frac{\beta_R\sigma_L(\gamma_R-\gamma_L)}{\beta_R\sigma_L+\beta_L\sqrt{\sigma_L\sigma_R}}\phi(\sigma_L,x,t)
+\frac{1}{2\pi}\int_{-\infty}^\infty e^{ikx-\edits{\lambda_L}t}\hat{v}_0^L(k)\ud k\\
&+\frac{\beta_R\sigma_L-\beta_L\sqrt{\sigma_L\sigma_R}}{2\pi(\beta_R \sigma_L+\beta_L\sqrt{\sigma_L\sigma_R})}\int_{\partial D^-} e^{ikx-\edits{\lambda_L}t}\hat{v}_0^L(-k)\ud k\\
&-\frac{\beta_R\sigma_L}{\pi(\beta_L\sigma_R+\beta_R\sqrt{\sigma_L\sigma_R})}\int_{\partial D^-} e^{ikx-\edits{\lambda_L}t}\hat{v}_0^R\left(k\sqrt{\frac{\sigma_L}{\sigma_R}}\right)\ud k,
\end{split}
\eeq
for $-\infty<x<0$ and for $0<x<\infty$,
\beq\label{2iqsolnsR}
\begin{split}
q^R(x,t)=&\gamma^R+\frac{\beta_L\sqrt{\sigma_L\sigma_R}(\gamma_R-\gamma_L)}{\beta_R\sigma_L+\beta_L\sqrt{\sigma_R\sigma_L}}\phi(\sigma_R,x,t)
+\frac{1}{2\pi}\int_{-\infty}^\infty e^{ikx-\edits{\lambda_R} t}\hat{v}_0^R(k)\ud k\\
&+\frac{\beta_L\sigma_R}{\pi\left(\beta_R\sigma_L+\beta_L\sqrt{\sigma_R\sigma_L}\right)}\int_{\partial D^+}e^{ikx-\edits{\lambda_R}t}\hat{v}_0^L\left(k\sqrt{\frac{\sigma_R}{\sigma_L}}\right)\ud k\\
&+\frac{\beta_R\sigma_L-\beta_L\sqrt{\sigma_L\sigma_R}}{2\pi\left(\beta_R\sigma_L+\beta_L\sqrt{\sigma_R\sigma_L}\right)}\int_{\partial D^+}e^{ikx-\edits{\lambda_R}t}\hat{v}_0^R(-k)\ud k.
\end{split}
\eeq
\end{prop}

\subsection{Remarks}
\begin{itemize}
\item The use of the discrete symmetries of the \edits{functions $\lambda_L$ and $\lambda_R$ or of the  dispersion relation} is an important aspect of the Fokas Method \cite{DeconinckTrogdonVasan, FokasBook, FokasPelloni4}.  When solving the LS equation in a single medium, the only discrete symmetry required is $k\to -k$, which was used here to obtain~\eqref{tglobal2iL minus}. Due to the two media, there are two \edits{functions} in the present problem: $\edits{\lambda_L}=-i\sigma_L k^2$ and $\edits{\lambda_R}=-i\sigma_R k^2$. The collection of both \edits{functions  $\{\lambda_L, \lambda_R\}$} retains the discrete symmetry $k\to -k$, but admits an additional one, namely: $k\to  k\sqrt{\sigma_R/\sigma_L}$ which transforms the two \edits{functions} to each other.  \textbf{All nontrivial discrete symmetries of $\{\edits{\lambda_L}, \edits{\lambda_R}\}$ are needed to derive the final solution representation}.

\item \edits{In equations~\eqref{2iqsolnsL} and~\eqref{2iqsolnsR} it is possible to deform the integration paths back to the real line. This deformation hints that a classical solution in terms of Fourier-like integral transforms should be possible. However, a priori it is not clear how to obtain the appropriate transforms for general initial conditions and boundary conditions. In effect, as in \cite{FokasBook}, the Fokas method can be seen as a method to construct the appropriate transform to solve the problem. }

\item It is interesting to note that when $\sigma_L=\sigma_R$, $\beta_L=\beta_R$ and $\gamma^L=\gamma^R=0$,  the solution formulae in their proper $x$-domain of definition reduce to the solution of the whole line problem.  Also, if $\gamma_L=0=\gamma_R$, Cascaval and Hunter~\cite{CascavalHunter} find a solution to the LS equation with an interface by imposing the solution for the LS problem on the half-line given in~\cite{FokasBook} and viewing the interface problem as a forced problem on the real line where the forcing is occurring at the interface.   This leads to a solution of the interface problem which requires the numerical solution of an integral equation.

\item The leading-order behavior in time for \eqref{2ieqns} with initial conditions which decay sufficiently fast to the boundary values \eqref{2i_bc} at $\pm \infty$ is easily obtained by using integration by parts and the method of stationary phase~\cite{BenderOrszag}.  In the limit as $t\to\infty$ for $x/t$ constant,
\beq\label{2i_Lsoln_LOB}
\begin{split}
q^L(x,t)&\sim\frac{\beta_R\gamma_R\sqrt{\sigma_L}+\beta_L\gamma_L\sqrt{\sigma_R}}{\beta_R\sqrt{\sigma_L}+\beta_L\sqrt{\sigma_R}}+\frac{e^{\frac{i\pi}{4}-\frac{ix^2}{4\sigma_Lt}}}{\sqrt{4\sigma_Lt}}\left(\frac{\beta_R\sigma_L(\gamma_R-\gamma_L)x}{\beta_R\sigma_L+\beta_L\sqrt{\sigma_L\sigma_R}}+\frac{\hat{v}_0^L\left(-\frac{x}{2\sigma_Lt}\right)}{\sqrt{\pi}}\right.\\
&\left.-\frac{(\beta_R\sigma_L-\beta_L\sqrt{\sigma_L\sigma_R})\hat{v}_0^L\left(\frac{x}{2\sigma_Lt}\right)}{(\beta_R \sigma_L+\beta_L\sqrt{\sigma_L\sigma_R})\sqrt{\pi}}
+\frac{\beta_R\sigma_L\hat{v}_0^R\left(-\frac{x}{2t\sqrt{\sigma_L\sigma_R}}\right)}{(\beta_L\sigma_R+\beta_R\sqrt{\sigma_L\sigma_R})\sqrt{\pi}} \right),
\end{split}
\eeq
for $-\infty<x<0$ and, for $0<x<\infty$,
\beq\label{2i_Rsoln_LOB}
\begin{split}
q^R(x,t)&\sim\frac{\beta_R\gamma_R\sqrt{\sigma_L}+\beta_L\gamma_L\sqrt{\sigma_R}}{\beta_R\sqrt{\sigma_L}+\beta_L\sqrt{\sigma_R}}+\frac{e^{\frac{i\pi}{4}-\frac{ix^2}{4\sigma_Rt}}}{\sqrt{4\sigma_Rt}}\left(\frac{\beta_L\sqrt{\sigma_L\sigma_R}(\gamma_R-\gamma_L)x}{\beta_R\sigma_L+\beta_L\sqrt{\sigma_R\sigma_L}}
+\frac{\hat{v}_0^R\left(-\frac{x}{2\sigma_Rt}\right)}{\sqrt{\pi}}\right.\\
&\left.+\frac{\beta_L\sigma_R\hat{v}_0^L\left(-\frac{x}{2t\sqrt{\sigma_R\sigma_L}}\right)}{\left(\beta_R\sigma_L+\beta_L\sqrt{\sigma_R\sigma_L}\right)\sqrt{\pi}}
+\frac{(\beta_R\sigma_L-\beta_L\sqrt{\sigma_L\sigma_R})\hat{v}_0^R\left(\frac{x}{2\sigma_Rt}\right)}{\left(\beta_R\sigma_L+\beta_L\sqrt{\sigma_R\sigma_L}\right)\sqrt{\pi}}\right).
\end{split}
\eeq

The constant factor in~\eqref{2i_Lsoln_LOB} and~\eqref{2i_Rsoln_LOB} is the weighted average of the boundary conditions at infinity with weights given by $\beta_R\sqrt{\sigma_L}$ and $\beta_L\sqrt{\sigma_R}$.  The oscillations are contained in the terms $\exp(-ix^2/(4\sigma_Lt))$ and $\exp(-ix^2/(4\sigma_Rt))$.  In Figure~\ref{fig:2i_LOB} the envelope of the real (imaginary) part of the solution is plotted in gray (black) as a dot-dashed line.  The real part of the solution (plotted as a solid line in blue) is centered around the weighted average (plotted in black as a dotted line) and the imaginary part of the solution (plotted as a dashed line in red) is centered around zero.  In using the method of stationary phase one must look in directions of constant $x/t$.  Using integration by parts there is no such restriction.  When $x$ is large the term from integration by parts is dominant and so in Figure~\ref{fig:2i_LOB} there is no need to fix $x/t$.

\begin{figure} 
   \centering
   \includegraphics[width=.7\textwidth]{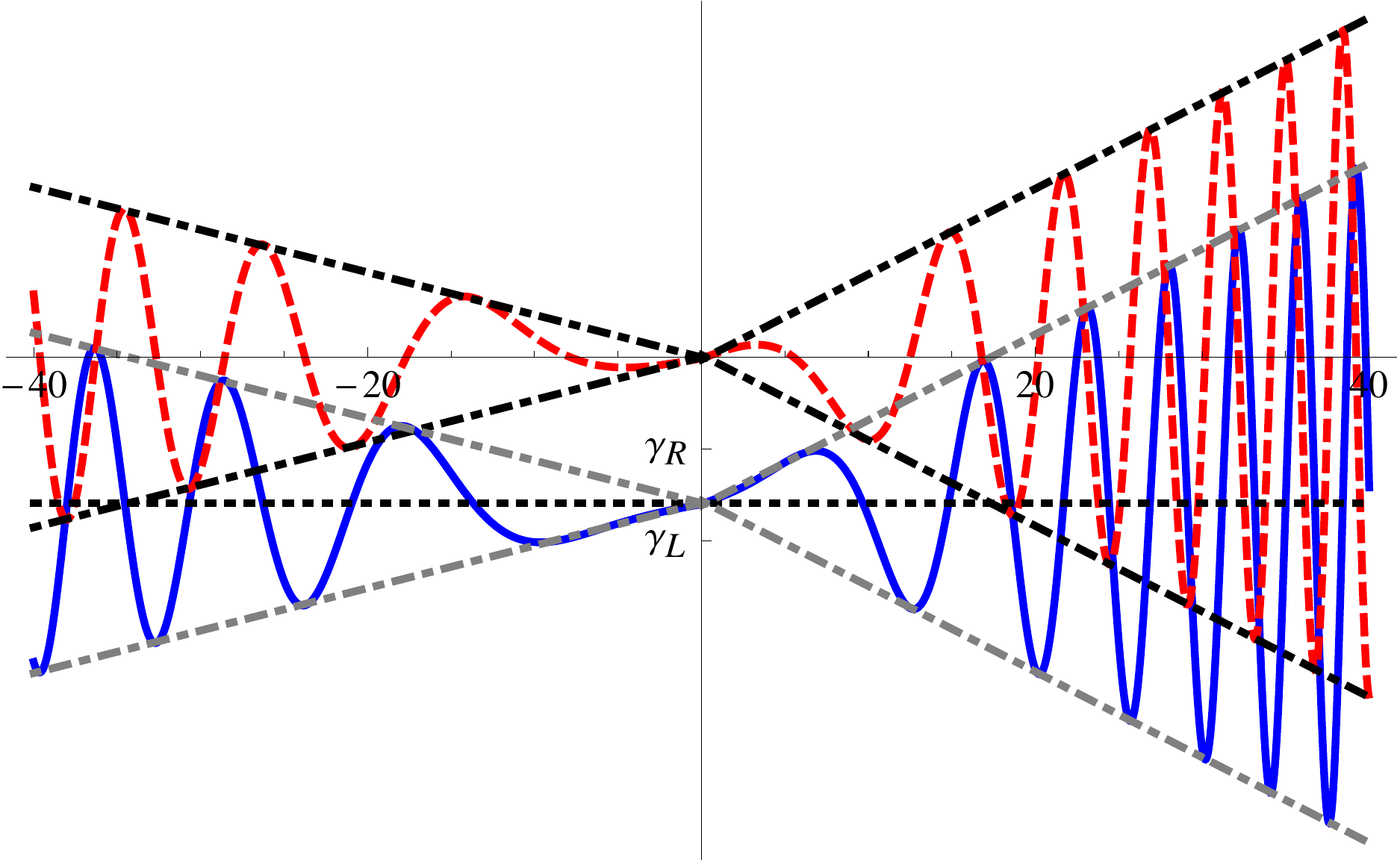} 
   \caption{The leading order behavior of $q(x,t)$ as given in~\eqref{2i_Lsoln_LOB} and~\eqref{2i_Rsoln_LOB} with $t=10, \gamma_L=-20, \gamma_R=-10, \beta_L=2, \beta_R=1, \sigma_L=2 $, and $\sigma_R=1$ and initial conditions $q_0^L(x)=\frac{\beta_L\gamma_L+\beta_R\gamma_R}{\beta_L+\beta_R}+\beta_R\frac{\gamma_R-\gamma_L}{\beta_L+\beta_R}\tanh(x)$ and $q_0^R(x)=\frac{\beta_L\gamma_L+\beta_R\gamma_R}{\beta_L+\beta_R}+\beta_L\frac{\gamma_R-\gamma_L}{\beta_L+\beta_R}\tanh(x)$.}
   \label{fig:2i_LOB}
\end{figure}

\item In quantum mechanics one considers only the finite energy case, that is, $\gamma_L=0=\gamma_R$.  In this case, asymptotics requires the use of the method of stationary phase only.  Thus, in Figure~\ref{fig:2i_LOB_qm} we consider solutions for $x/t=\pm1$.  The real and imaginary parts of the solution are centered around 0.  In Figure~\ref{fig:2i_LOB_qm} the real part of the solution is plotted as a solid line in blue and the imaginary part of the solution is plotted as a dashed line in red.  The envelope of the real (imaginary) part of the solution is plotted in gray (black) as a dot-dashed line.

\begin{figure} 
   \centering
   \includegraphics[width=.4\textwidth]{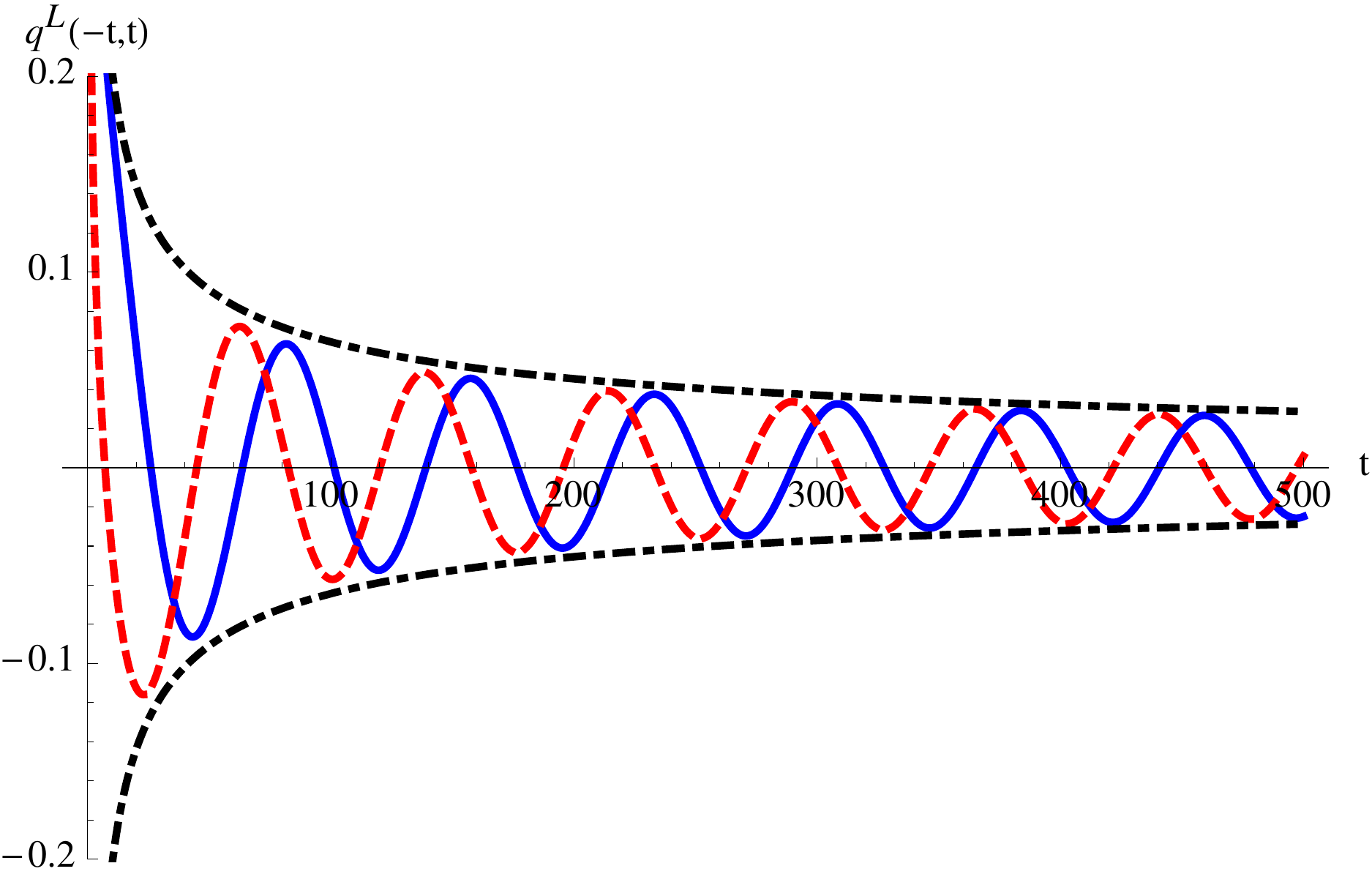} 
      \includegraphics[width=.4\textwidth]{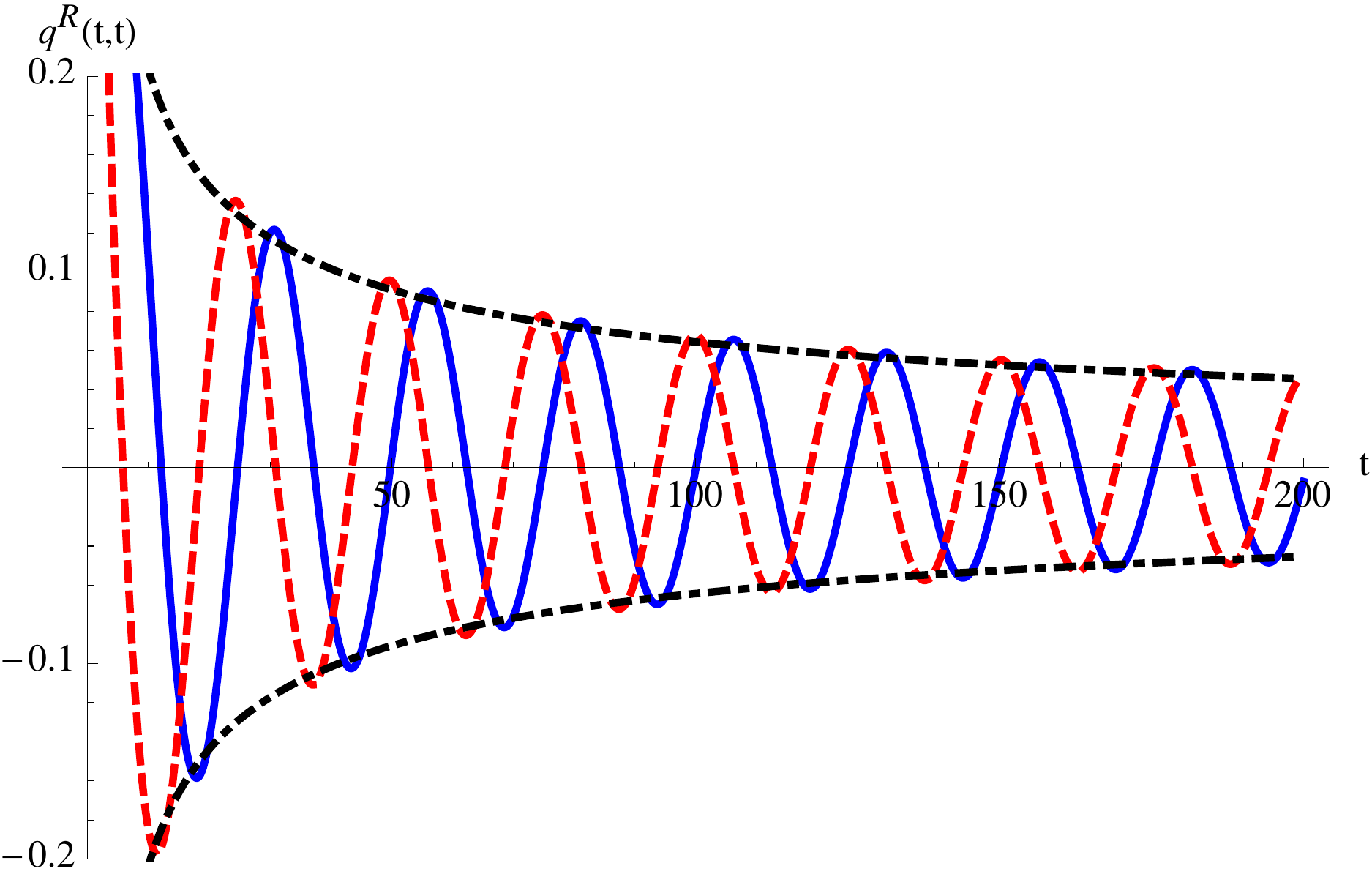} 
   \caption{The leading order behavior of $q^L(-t,t)$ and $q^R(t,t)$ as given in~\eqref{2i_Lsoln_LOB} and~\eqref{2i_Rsoln_LOB} respectively with $\gamma_L=0, \gamma_R=0,\beta_L=4, \beta_R=1, \sigma_L=3$, and $\sigma_R=1$ with $q_0^L(x)=(1+\beta_Rx)e^{-x^2}$ and $q_0^R(x)=(1+\beta_Lx)e^{-x^2}$.}
   \label{fig:2i_LOB_qm}
\end{figure}
\end{itemize}

\section{Two finite domains}\label{sec:ff}

We wish to find $q^L(x,t)$ and $q^R(x,t)$ satisfying
\beq\label{2feqns}
\begin{aligned}
i q^L_t(x,t)=&&\sigma_Lq^L_{xx}(x,t)&~~~~&-a<&x<0,&t>0,\\
i q^R_t(x,t)=&&\sigma_Rq^R_{xx}(x,t)&~~~~&0<&x<b,&t>0,
\end{aligned}
\eeq
subject to the Robin boundary conditions 
\beq\label{2fbcs}
\begin{aligned}
\alpha_1 q^L(-a,t)+\alpha_2 q^L_x(-a,t)=&f^L(t),~~&t>0,\\
\alpha_3 q^R(b,t)+\alpha_4 q^R_x(b,t)=&f^R(t),~~&t>0,
\end{aligned}
\eeq
the initial conditions
\beq
\begin{aligned}
q^L(x,0)=&q^L_0(x),~~~&-a<&x<0,\\
q^R(x,0)=&q^R_0(x),~~~&0<&x<b,
\end{aligned}
\eeq
and the interface conditions 
\beq\label{2f_ifc}
\begin{aligned}
q^L(0,t)=&q^R(0,t),~~&t>0,\\
\beta_L q^L_x(0,t)=&\beta_R q^R_x(0,t),~~&t>0,
\end{aligned}
\eeq
where  $a>0$, $b>0$, and $\beta_L$, $\beta_R$, and $\alpha_i$, $1\leq i\leq4$ are nonzero, $t$-independent constants. As before we assume $\sigma_L$ and $\sigma_R$ are positive for convenience.  If $\alpha_1=\alpha_3=0$ then Neumann boundary conditions are prescribed, whereas if $\alpha_2=\alpha_4=0$ then Dirichlet conditions are given.

As before we begin with the local relations

\begin{subequations}
\begin{align}
\label{local2fL}
\left(e^{-ikx+\edits{\lambda_L}t}q^L(x,t)\right)_t=&\left(\sigma_Le^{-ikx+\edits{\lambda_L}t}(kq^L(x,t)-iq^L_x(x,t))\right)_x,\\
\label{local2fR}
\left(e^{-ikx+\edits{\lambda_R}t}q^L(x,t)\right)_t=&\left(\sigma_Re^{-ikx+\edits{\lambda_R}t}(kq^R(x,t)-iq^R_x(x,t))\right)_x.
\end{align}
\end{subequations}
For $k\in\CC$ we define the following transforms:

\begin{align*}
\hat{f}_L(\omega,t)=&\int_0^t e^{\omega s}f_L(s)\ud s, &\hat{f}_R(\omega,t)=&\int_0^t e^{\omega s}f_R(s)\ud s,\\
h_1^L(\omega,t)=&\int_0^te^{\omega s}q^L_x(-a,s)\ud s, & h_0^L(\omega,t)=&\int_0^t e^{\omega s} q^L(-a,s)\ud s,\\
h_1^R(\omega,t)=&\int_0^te^{\omega s}q^R_x(b,s)\ud s, & h_0^R(\omega,t)=&\int_0^t e^{\omega s} q^R(b,s)\ud s,\\
\hat{q}^L(k,t)=&\int_{-a}^0e^{-ikx}q^L(x,t)\ud x, &\hat{q}^L_0(k)=&\int_{-a}^0e^{-ikx}q^L_0(x)\ud x,\\
\hat{q}^R(k,t)=&\int_{0}^b e^{-ikx}q^R(x,t)\ud x, &\hat{q}^R_0(k)=&\int_{0}^b e^{-ikx}q^R_0(x)\ud x,\\
g_{0}({\omega},t)=&\int_{0}^te^{\omega s}q^L(0,s)\ud s=\int_{0}^te^{\omega s}q^R(0,s)\ud s,\\
g_{1}({\omega},t)=&\int_{0}^te^{{\omega} s}q^L_x(0,s)\ud s=\frac{\beta_R}{\beta_L} \int_{0}^te^{{\omega} s}q^R_x(0,s)\ud s.
\end{align*}

\begin{figure}[htbp!] 
   \centering
   \def\svgwidth{4in}
   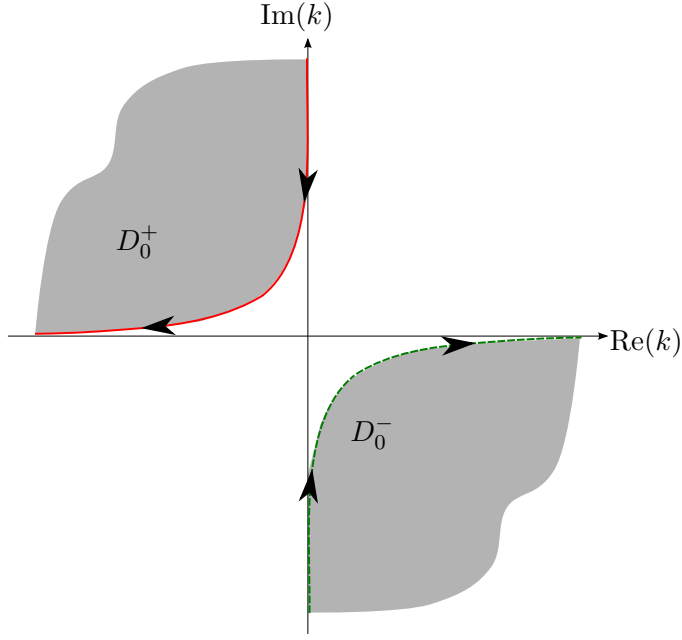
      \caption{Deformation of the contours in Figure~\ref{fig:LS} away from the real axis.
   \label{fig:LS_nonzero}}
\end{figure}

Applying Green's Theorem~\cite{AblowitzFokas} in the domains $[-a,0]\times [0,t]$ and $[0,b]\times [0,t]$ respectively, we find the global relations

\begin{subequations}\label{global2f}
\begin{align}
\label{global2fL}
e^{\edits{\lambda_L} t}\hat{q}^L(k,t)=&\hat{q}_0^L(k)+k\sigma_Lg_0(\edits{\lambda_L},t)-i\sigma_Lg_1(\edits{\lambda_L},t)-\sigma_Le^{ika}(kh_0^L(\edits{\lambda_L},t)-ih_1^L(\edits{\lambda_L},t)) ,\\
\label{global2fR}
e^{\edits{\lambda_R}t}\hat{q}^R(k,t)=&\hat{q}_0^R(k)-k\sigma_Rg_0(\edits{\lambda_R},t) +\frac{i\sigma_R\beta_L}{\beta_R}g_1(\edits{\lambda_R},t)+\sigma_Re^{-ikb}(kh_0^R(\edits{\lambda_R},t)-ih_1^R(\edits{\lambda_R},t)),
\end{align}
\end{subequations}
which are valid for all $k\in\CC$ in contrast to~\eqref{global2i}.  Using the invariance of $\edits{\lambda_L}(k)$ and $\edits{\lambda_R}(k)$ under $k\to-k$ we supplement~\eqref{global2f} with their evaluation at $-k$, namely
\begin{subequations}\label{global2f minus}
\begin{align}
\label{global2fL minus}
e^{\edits{\lambda_L} t}\hat{q}^L(-k,t)=&\hat{q}_0^L(-k)-k\sigma_Lg_0(\edits{\lambda_L},t)-i\sigma_Lg_1(\edits{\lambda_L},t)-\sigma_Le^{-ika}(-kh_0^L(\edits{\lambda_L},t)-ih_1^L(\edits{\lambda_L},t)) ,\\
\label{global2fR minus}
e^{\edits{\lambda_R}t}\hat{q}^R(-k,t)=&\hat{q}_0^R(-k)+k\sigma_Rg_0(\edits{\lambda_R},t) +\frac{i\sigma_R\beta_L}{\beta_R}g_1(\edits{\lambda_R},t)+\sigma_Re^{ikb}(-kh_0^R(\edits{\lambda_R},t)-ih_1^R(\edits{\lambda_R},t)),
\end{align}
\end{subequations}
Inverting the Fourier transform in~\eqref{global2fL} we have
\bea
q^L(x,t)=&&\frac{1}{2\pi}\int_{-\infty}^\infty e^{ikx-\edits{\lambda_L}t}\hat{q}_0^L(k)\ud k+\frac{1}{2\pi}\int_{-\infty}^\infty e^{ikx-\edits{\lambda_L}t}\sigma_L(kg_0(\edits{\lambda_L},t)-ig_1(\edits{\lambda_L},t))\ud k\\
&&+\frac{1}{2\pi}\int_{-\infty}^\infty \sigma_Le^{ik(x+a)-\edits{\lambda_L}t}(kh_0^L(\edits{\lambda_L},t)-ih_1^L(\edits{\lambda_L},t))\ud k,
\eea
for $-a<x<0$ and $t>0$.  The integrand of the second integral is entire and decays as $k\to\infty$ for $k\in \CC^-\setminus D^-$.  The last integral is entire and decays as $k\to\infty$ for $k\in\CC^+\setminus D^+$.  It is convenient to deform both contours away from the real axis to avoid singularities in the integrands that become apparent in what follows.  Initially these singularities are removable since the integrands are entire.  Writing integrals of sums as sums of integrals, these singularities may cease to be removable.  With the deformation away from the real axis the singularities are no cause for concern.  In other words, we deform $D^+$ to $D^+_0$ and $D^-$ to $D^-_0$ as show in Figure~\ref{fig:LS_nonzero} where the deformed contours approach the real axis asymptotically. Thus,
\beq\label{soln2fL}
\begin{split}
q^L(x,t)=&\frac{1}{2\pi}\int_{-\infty}^\infty e^{ikx-\edits{\lambda_L}t}\hat{q}_0^L(k)\ud k-\frac{1}{2\pi}\int_{\partial D_0^-} e^{ikx-\edits{\lambda_L}t}\sigma_L(kg_0(\edits{\lambda_L},t)-ig_1(\edits{\lambda_L},t))\ud k\\
&+\frac{1}{2\pi}\int_{\partial D_0^+} \sigma_Le^{ik(x+a)-\edits{\lambda_L}t}(kh_0^L(\edits{\lambda_L},t)-ih_1^L(\edits{\lambda_L},t))\ud k.
\end{split}
\eeq

\no Similarly, inverting the Fourier transform in~\eqref{global2fR} we have
\begin{equation*}
\begin{split}
q^R(x,t)=&\frac{1}{2\pi}\int_{-\infty}^\infty e^{ikx-\edits{\lambda_R} t}\hat{q}_0^R(k)\ud k+\frac{1}{2\pi}\int_{-\infty}^\infty e^{ikx-\edits{\lambda_R} t}\sigma_R\left(-kg_0(\edits{\lambda_R},t)+\frac{i\beta_L}{\beta_R}g_1(\edits{\lambda_R},t)\right)\ud k\\
&+\frac{1}{2\pi}\int_{-\infty}^\infty e^{ik(x-b)-\edits{\lambda_R} t}\sigma_R(kh_0^R(\edits{\lambda_R},t)-ih_1^R(\edits{\lambda_R},t))\ud k,
\end{split}
\end{equation*}

\no for $0<x<b$ and $t>0$. The integrand of the second integral is entire and decays as $k\to\infty$ for $k\in \CC^+\setminus D^+$. The integrand of the third integral is entire and decays as $k\to\infty$ for $k\in\CC^-\setminus D^-$. We deform as above to find
\beq\label{soln2fR}
\begin{split}
q^R(x,t)=&\frac{1}{2\pi}\int_{-\infty}^\infty e^{ikx-\edits{\lambda_R} t}\hat{q}_0^R(k)\ud k
+\frac{1}{2\pi}\int_{\partial D_0^+} e^{ikx-\edits{\lambda_R} t}\sigma_R\left(-kg_0(\edits{\lambda_R},t)+\frac{i\beta_L}{\beta_R}g_1(\edits{\lambda_R},t)\right)\ud k\\
&-\frac{1}{2\pi}\int_{\partial D_0^-} e^{ik(x-b)-\edits{\lambda_R} t}\sigma_R(kh_0^R(\edits{\lambda_R},t)-ih_1^R(\edits{\lambda_R},t))\ud k.\end{split}
\eeq

Taking the time transform of the boundary conditions results in

\beq\label{bcL_t}
\alpha_1h_0^L(\omega,t)+\alpha_2h_1^L(\omega,t)=\hat{f}^L(\omega,t),
\eeq

\no and
\beq\label{bcR_t}
\alpha_3h_0^R(\omega,t)+\alpha_4h_1^R(\omega,t)=\hat{f}^R(\omega,t).
\eeq

To obtain a system of six equations for the six unknown functions $g_0(\omega,t)$, $g_1(\omega,t)$, $h_0^L(\omega,t)$, $h_1^L(\omega,t)$, $h_0^R(\omega,t)$, and $h_1^R(\omega,t)$ we use the  global relations evaluated at $k$ and $-k$~\eqref{global2f} and~\eqref{global2f minus} and the time transform of the boundary conditions~\eqref{bcL_t} and~\eqref{bcR_t}.  

Although we could solve this problem in its full generality, we restrict to the case of Dirichlet boundary conditions ($\alpha_2=\alpha_4=0$), to simplify the already cumbersome formulae below. The system is not solvable for $h_1^L(\omega, t)$ and $h_1^R(\omega,t)$ if $\Delta_L(k)=0$, where

\begingroup
\setlength{\thinmuskip}{0mu}
\setlength{\medmuskip}{0mu}
\setlength{\thickmuskip}{0mu}
\beq\label{deltaL}
\Delta_L(k)
=4i\pi\alpha_1\alpha_3\sigma_Re^{ik\left(a+b\sqrt{\frac{\sigma_L}{\sigma_R}}\right)}\left(\beta_L\cos(ak)\sin\left(bk\sqrt{\frac{\sigma_L}{\sigma_R}}\right)+\beta_R\sqrt{\frac{\sigma_L}{\sigma_R}}\sin(ak) \cos\left(bk\sqrt{\frac{\sigma_L}{\sigma_R}}\right) \right).
\eeq
\endgroup

\no It is easily seen that all values of $k$ satisfying this relation (including $k=0$) are on the real line. Thus on the contours, the equations are solved without problem, resulting in the expressions below. As in the previous section, the right-hand sides of these expressions involve $\hat q^L(k,t)$ and $\hat q^R(k,t)$, evaluated at a variety of arguments. All terms with such dependence are written out explicitly below. Terms that depend on known quantities only are contained in $K^L(x,t)$ and $K^R(x,t)$, the expressions for which are given in Proposition~\ref{prop:2f}.
\begingroup
\setlength{\thinmuskip}{0mu}
\setlength{\medmuskip}{0mu}
\setlength{\thickmuskip}{0mu}
\beq\label{soln2fLfull}
\begin{split}
q^L(x,t)=&K^L(x,t)-\int_{\partial D_0^-}  \frac{\alpha_1\alpha_3\sigma_R}{2\Delta_L(k)} e^{ikx}\left(\beta_L \left(e^{2ibk\sqrt{\frac{\sigma_L}{\sigma_R}}}-1\right) -\beta_R\sqrt{\frac{\sigma_L}{\sigma_R}}\left(e^{2ibk\sqrt{\frac{\sigma_L}{\sigma_R}}}+1\right)  \right)\hat{q}^L(k,t) \ud k \\
&+\int_{\partial D_0^-} \frac{\alpha_1\alpha_3\sigma_R}{2\Delta_L(k)} e^{ik(2a+x)}\left(\beta_L \left(e^{2ibk\sqrt{\frac{\sigma_L}{\sigma_R}}}-1\right) -\beta_R\sqrt{\frac{\sigma_L}{\sigma_R}}\left(e^{2ibk\sqrt{\frac{\sigma_L}{\sigma_R}}}+1\right)  \right) \hat{q}^L(-k,t) \ud k \\
&+\int_{\partial D_0^-}\frac{\alpha_1\alpha_3\beta_R\sigma_L}{\Delta_L(k)}e^{ik(2a+2b\sqrt{\frac{\sigma_L}{\sigma_R}}+x)}  \hat{q}^R\left(k\sqrt{\frac{\sigma_L}{\sigma_R}},t\right) \ud k \\
&-\int_{\partial D_0^-}  \frac{\alpha_1\alpha_3\beta_R\sigma_L}{\Delta_L(k)} e^{ik(2a+x)}\hat{q}^R\left(-k\sqrt{\frac{\sigma_L}{\sigma_R}},t\right) \ud k \\
&-\int_{\partial D_0^+}\frac{\alpha_1\alpha_3\sigma_R}{2\Delta_L(k)} e^{ik(2a+x)}\left( \beta_L\left(e^{2ibk\sqrt{\frac{\sigma_L}{\sigma_R}}}-1\right)+\beta_R\sqrt{\frac{\sigma_L}{\sigma_R}}\beta_R \left(e^{2ibk\sqrt{\frac{\sigma_L}{\sigma_R}}}+1\right) \right)   \hat{q}^L(k,t) \ud k\\
&-\int_{\partial D_0^+}\frac{\alpha_1\alpha_3\sigma_R}{2\Delta_L(k)} e^{ik(2a+x)}\left( \beta_L\left(e^{2ibk\sqrt{\frac{\sigma_L}{\sigma_R}}}-1\right)-\beta_R\sqrt{\frac{\sigma_L}{\sigma_R}}\beta_R \left(e^{2ibk\sqrt{\frac{\sigma_L}{\sigma_R}}}+1\right) \right)   \hat{q}^L(-k,t) \ud k\\
&-\int_{\partial D_0^+}\frac{\alpha_1\alpha_3\beta_R\sigma_L}{\Delta_L(k)} e^{ik(2a+2b\sqrt{\frac{\sigma_L}{\sigma_R}}+x)}  \hat{q}^R\left(k\sqrt{\frac{\sigma_L}{\sigma_R}},t\right) \ud k\\
&+\int_{\partial D_0^+}\frac{\alpha_1\alpha_3\beta_R\sigma_L}{\Delta_L(k)}e^{ik(2a+x)}   \hat{q}^R\left(-k\sqrt{\frac{\sigma_L}{\sigma_R}},t\right) \ud k,
\end{split}
\eeq
\endgroup

\no for $-a<x<0$, $t>0$ and
\begingroup
\setlength{\thinmuskip}{0mu}
\setlength{\medmuskip}{0mu}
\setlength{\thickmuskip}{0mu}
 \beq\label{soln2fRfull}
\begin{split}
q^R(x,t)=&K^R(x,t)+\int_{\partial D_0^+}  \frac{\alpha_1\alpha_3\beta_L\sigma_R}{\Delta_R(k)} e^{ikx}\hat{q}^L\left(k\sqrt{\frac{\sigma_R}{\sigma_L}},t\right) \ud k \\
&-\int_{\partial D_0^+} \frac{\alpha_1\alpha_3\beta_L\sigma_R}{\Delta_R(k)} e^{ik(2a\sqrt{\frac{\sigma_R}{\sigma_L}}+x)} \hat{q}^L\left(-k\sqrt{\frac{\sigma_R}{\sigma_L}},t\right) \ud k \\
&+\int_{\partial D_0^+}\frac{\alpha_1\alpha_3\sigma_L}{2\Delta_R(k)}e^{ik(2b+x)}\left(\beta_R\left(e^{2iak\sqrt{\frac{\sigma_R}{\sigma_L}}}-1\right)+\beta_R\sqrt{\frac{\sigma_R}{\sigma_L}}\left(e^{2iak\sqrt{\frac{\sigma_R}{\sigma_L}}}+1\right)  \right)  \hat{q}^R(k,t) \ud k \\
&-\int_{\partial D_0^+}  \frac{\alpha_1\alpha_3\sigma_L}{2\Delta_R(k)} e^{ikx}\left(\beta_R\left(e^{2iak\sqrt{\frac{\sigma_R}{\sigma_L}}}-1\right)+\beta_L\sqrt{\frac{\sigma_R}{\sigma_L}}\left(e^{2iak\sqrt{\frac{\sigma_R}{\sigma_L}}}+1\right)  \right)    \hat{q}^R(-k,t) \ud k \\
&-\int_{\partial D_0^-}\frac{\alpha_1\alpha_3\beta_L\sigma_R}{\Delta_R(k)} e^{ikx}  \hat{q}^L\left(k\sqrt{\frac{\sigma_R}{\sigma_L}},t\right) \ud k\\
&+\int_{\partial D_0^-}\frac{\alpha_1\alpha_3\beta_L\sigma_R}{\Delta_R(k)} e^{ik(2a\sqrt{\frac{\sigma_R}{\sigma_L}}+x)}  \hat{q}^L\left(-k\sqrt{\frac{\sigma_R}{\sigma_L}},t\right) \ud k\\
&+\int_{\partial D_0^-}\frac{\alpha_1\alpha_3\sigma_L}{2\Delta_R(k)} e^{ikx}\left(\beta_R\left(e^{2iak\sqrt{\frac{\sigma_R}{\sigma_L}}}-1\right)-\beta_L\sqrt{\frac{\sigma_R}{\sigma_L}}\left(e^{2iak\sqrt{\frac{\sigma_R}{\sigma_L}}}+1\right)  \right)  \hat{q}^R(k,t) \ud k\\
&+\int_{\partial D_0^-}\frac{\alpha_1\alpha_3\sigma_L}{2\Delta_R(k)}e^{ikx}\left(\beta_R\left(e^{2iak\sqrt{\frac{\sigma_R}{\sigma_L}}}-1\right)+\beta_L\sqrt{\frac{\sigma_R}{\sigma_L}}\left(e^{2iak\sqrt{\frac{\sigma_R}{\sigma_L}}}+1\right)  \right)   \hat{q}^R(-k,t) \ud k,
\end{split}
\eeq
\endgroup
 
\no for $0<x<b$, $t>0$, where $$\Delta_R(k)=\sqrt{\frac{\sigma_R}{\sigma_L}}\Delta_L\left(k\sqrt{\frac{\sigma_R}{\sigma_L}}\right).$$

The integrands written explicitly in~\eqref{soln2fLfull} and~\eqref{soln2fRfull} decay in the regions around whose boundaries they are integrated.  Thus, using Jordan's Lemma and Cauchy's Theorem these integrals are shown to vanish.  Thus the final solution is given by $K^L(x,t)$ and $K^R(x,t)$.

\begin{prop}\label{prop:2f} The solution of the linear Schr\"odinger interface problem~\eqref{2feqns}-\eqref{2f_ifc} is given by
\begin{align}
\begin{split}
q^L(x,t)=&K^L(x,t)=\frac{1}{2\pi}\int_{-\infty}^\infty e^{ikx+ik^2\sigma_Lt}\hat{q}_0^L(k)\ud k\\
&+\int_{\partial D_0^-} \frac{\alpha_1\alpha_3\sigma_R}{2\Delta_L(k)}e^{ikx+ik^2\sigma_L t}\left( \beta_L\left(e^{2ibk\sqrt{\frac{\sigma_L}{\sigma_R}}}-1\right)-\beta_R\sqrt{\frac{\sigma_L}{\sigma_R}}\left(e^{2ibk\sqrt{\frac{\sigma_L}{\sigma_R}}}+1\right) \right)  \hat{q}_0^L(k) \ud k\\
&-\int_{\partial D_0^-} \frac{\alpha_1\alpha_3\sigma_R}{2\Delta_L(k)} e^{ik(2a+x)+ik^2\sigma_L t}\left( \beta_L\left(e^{2ibk\sqrt{\frac{\sigma_L}{\sigma_R}}}-1\right)-\beta_R\sqrt{\frac{\sigma_L}{\sigma_R}}\left(e^{2ibk\sqrt{\frac{\sigma_L}{\sigma_R}}}+1\right) \right)  \hat{q}_0^L(-k) \ud k\\
&-\int_{\partial D_0^-} \frac{\alpha_1\alpha_3\beta_R\sigma_L}{\Delta_L(k)}e^{ik(2a+2b\sqrt{\frac{\sigma_L}{\sigma_R}}+x)+ik^2\sigma_Lt}  \hat{q}_0^R\left(k\sqrt{\frac{\sigma_L}{\sigma_R}}\right) \ud k\\
&+\int_{\partial D_0^-} \frac{\alpha_1\alpha_3\beta_R\sigma_L}{\Delta_L(k)}e^{ik(2a+x)+ik^2\sigma_Lt}   \hat{q}_0^L\left(-k\sqrt{\frac{\sigma_L}{\sigma_R}}\right) \ud k\\
&-\int_{\partial D_0^-} \frac{k\alpha_3\sigma_L\sigma_R}{\Delta_L(k)}e^{ik(a+x)+ik^2\sigma_Lt}\left(\beta_L\left(e^{2ibk\sqrt{\frac{\sigma_L}{\sigma_R}}}-1\right)-\beta_R\sqrt{\frac{\sigma_L}{\sigma_R}}\left(e^{2ibk\sqrt{\frac{\sigma_L}{\sigma_R}}}+1\right)   \right)  \hat{f}_L(\edits{\lambda_L},t) \ud k\\
&-\int_{\partial D_0^-} \frac{2k\alpha_1\beta_R\sigma_L\sqrt{\sigma_L\sigma_R}}{\Delta_L(k)}e^{ik(2a+b\sqrt{\frac{\sigma_L}{\sigma_R}}+x)+ik^2\sigma_Lt}  \hat{f}_R(\edits{\lambda_R},t) \ud k\\
&+\int_{\partial D_0^+} \frac{\alpha_1\alpha_3\sigma_R}{2\Delta_L(k)}e^{ik(2a+x)+ik^2\sigma_Lt}\left( \beta_L\left(e^{2ibk\sqrt{\frac{\sigma_L}{\sigma_R}}}-1\right)+\beta_R\sqrt{\frac{\sigma_L}{\sigma_R}}\left(e^{2ibk\sqrt{\frac{\sigma_L}{\sigma_R}}}+1\right) \right)  \hat{q}_0^L(k) \ud k\\
&+\int_{\partial D_0^+}  \frac{\alpha_1\alpha_3\sigma_R}{2\Delta_L(k)}e^{ik(2a+x)+ik^2\sigma_Lt}\left( \beta_L\left(e^{2ibk\sqrt{\frac{\sigma_L}{\sigma_R}}}-1\right)-\beta_R\sqrt{\frac{\sigma_L}{\sigma_R}}\left(e^{2ibk\sqrt{\frac{\sigma_L}{\sigma_R}}}+1\right) \right)  \hat{q}_0^L(-k) \ud k\\
&+\int_{\partial D_0^+} \frac{\alpha_1\alpha_3\beta_R\sigma_L}{\Delta_L(k)} e^{ik(2a+2b\sqrt{\frac{\sigma_L}{\sigma_R}}+x)+ik^2\sigma_Lt} \hat{q}_0^R\left(k\sqrt{\frac{\sigma_L}{\sigma_R}}\right) \ud k\\
&-\int_{\partial D_0^+} \frac{\alpha_1\alpha_3\beta_R\sigma_L}{\Delta_L(k)} e^{ik(2a+x)+ik^2\sigma_Lt} \hat{q}_0^L\left(-k\sqrt{\frac{\sigma_L}{\sigma_R}}\right) \ud k\\
&+\int_{\partial D_0^+} \frac{k\alpha_3\sigma_L\sigma_R}{\Delta_L(k)}e^{ik(a+x)+ik^2\sigma_Lt}\left(\beta_L\left(e^{2ibk\sqrt{\frac{\sigma_L}{\sigma_R}}}-1\right)-\beta_R\sqrt{\frac{\sigma_L}{\sigma_R}}\left(e^{2ibk\sqrt{\frac{\sigma_L}{\sigma_R}}}+1\right)  \right)  \hat{f}_L(\edits{\lambda_L},t) \ud k\\
&+\int_{\partial D_0^+} \frac{2k\alpha_1\beta_R\sigma_L\sqrt{\sigma_L\sigma_R}}{\Delta_L(k)}e^{ik(2a+b\sqrt{\frac{\sigma_L}{\sigma_R}}+x)+ik^2\sigma_Lt}  \hat{f}_R(\edits{\lambda_R},t) \ud k,
\end{split}
\end{align}

\no for $-a<x<0$, and, for $0<x<b$

\begin{align}
\begin{split}
q^R(x,t)=&K^R=\frac{1}{2\pi}\int_{-\infty}^\infty e^{ikx+ik^2\sigma_R t}\hat{q}_0^R(k)\ud k\\
&-\int_{\partial D_0^+} \frac{\alpha_1\alpha_3\beta_L\sigma_R}{\Delta_R(k)}e^{ikx+ik^2\sigma_Rt} \hat{q}_0^L\left(k\sqrt{\frac{\sigma_R}{\sigma_L}}\right) \ud k\\
&+\int_{\partial D_0^+} \frac{\alpha_1\alpha_3\beta_L\sigma_R}{\Delta_R(k)}e^{ik(2a\sqrt{\frac{\sigma_R}{\sigma_L}}+x)+ik^2\sigma_Rt} \hat{q}_0^L\left(-k\sqrt{\frac{\sigma_R}{\sigma_L}}\right) \ud k\\
&-\int_{\partial D_0^+} \frac{\alpha_1\alpha_3\sigma_L}{2\Delta_R(k)}e^{ik(2b+x)+ik^2\sigma_Rt}\left(\beta_R\left(e^{2iak\sqrt{\frac{\sigma_R}{\sigma_L}}}-1\right) +\beta_L\sqrt{\frac{\sigma_R}{\sigma_L}}\left(e^{2iak\sqrt{\frac{\sigma_R}{\sigma_L}}}+1\right)  \right) \hat{q}_0^R(k) \ud k\\
&+\int_{\partial D_0^+} \frac{\alpha_1\alpha_3\sigma_L}{2\Delta_R(k)}e^{ikx+ik^2\sigma_Rt}\left(\beta_R\left(e^{2iak\sqrt{\frac{\sigma_R}{\sigma_L}}}-1\right) +\beta_L\sqrt{\frac{\sigma_R}{\sigma_L}}\left(e^{2iak\sqrt{\frac{\sigma_R}{\sigma_L}}}+1\right)  \right) \hat{q}_0^R(-k) \ud k\\
&+\int_{\partial D_0^+} \frac{2k\alpha_3\beta_L\sigma_R\sqrt{\sigma_L\sigma_R}}{\Delta_R(k)}e^{ik(a\sqrt{\frac{\sigma_R}{\sigma_L}}+x)+ik^2\sigma_Rt} \hat{f}_L(\edits{\lambda_R},t) \ud k\\
&-\int_{\partial D_0^+} \frac{k\alpha_1\sigma_L\sigma_R}{\Delta_R(k)}e^{ik(b+x)+ik^2\sigma_Rt}\left(\beta_R\left(e^{2iak\sqrt{\frac{\sigma_R}{\sigma_L}}}-1\right) +\beta_L\sqrt{\frac{\sigma_R}{\sigma_L}}\left(e^{2iak\sqrt{\frac{\sigma_R}{\sigma_L}}}+1\right)  \right)  \hat{f}_R(\edits{\lambda_R},t) \ud k\\
&+\int_{\partial D_0^-} \frac{\alpha_1\alpha_3\beta_L\sigma_R}{\Delta_R(k)}e^{ikx+ik^2\sigma_Rt} \hat{q}_0^L\left(k\sqrt{\frac{\sigma_R}{\sigma_L}}\right) \ud k\\
&-\int_{\partial D_0^-} \frac{\alpha_1\alpha_3\beta_L\sigma_R}{\Delta_R(k)}e^{ik(2a\sqrt{\frac{\sigma_R}{\sigma_L}}+x)+ik^2\sigma_Rt} \hat{q}_0^L\left(-k\sqrt{\frac{\sigma_R}{\sigma_L}}\right) \ud k\\
&-\int_{\partial D_0^-} \frac{\alpha_1\alpha_3\sigma_L}{2\Delta_R(k)}e^{ikx+ik^2\sigma_Rt} \left( \beta_R\left(e^{2iak\sqrt{\frac{\sigma_R}{\sigma_L}}}-1\right)-\beta_L\sqrt{\frac{\sigma_R}{\sigma_L}}\left(e^{2iak\sqrt{\frac{\sigma_R}{\sigma_L}}}+1\right) \right)\hat{q}_0^R(k) \ud k\\
&-\int_{\partial D_0^-} \frac{\alpha_1\alpha_3\sigma_L}{2\Delta_R(k)}e^{ikx+ik^2\sigma_Rt} \left( \beta_R\left(e^{2iak\sqrt{\frac{\sigma_R}{\sigma_L}}}-1\right)+\beta_L\sqrt{\frac{\sigma_R}{\sigma_L}}\left(e^{2iak\sqrt{\frac{\sigma_R}{\sigma_L}}}+1\right) \right)\hat{q}_0^R(-k) \ud k\\
&-\int_{\partial D_0^-} \frac{2k\alpha_3\beta_L\sigma_R\sqrt{\sigma_L\sigma_R}}{\Delta_R(k)}e^{ik(a\sqrt{\frac{\sigma_R}{\sigma_L}}+x)+ik^2\sigma_Rt} \hat{f}_L(\edits{\lambda_R},t) \ud k\\
&+\int_{\partial D_0^-} \frac{k\alpha_1\sigma_L\sigma_R}{\Delta_R(k)}e^{ik(x+b)+ik^2\sigma_Rt} \left( \beta_R\left(e^{2iak\sqrt{\frac{\sigma_R}{\sigma_L}}}-1\right)+\beta_L\sqrt{\frac{\sigma_R}{\sigma_L}}\left(e^{2iak\sqrt{\frac{\sigma_R}{\sigma_L}}}+1\right) \right) \hat{f}_R(\edits{\lambda_R},t) \ud k.
\end{split}
\end{align}

\end{prop}

\subsection{Remarks}

\begin{itemize}

\item The solution of the problem posed in~\eqref{2feqns}-\eqref{2f_ifc} may be obtained using the classical method of separation of variables and superposition as was done for the heat equation in~\cite{HahnO}.  The solutions $q^L(x,t)$ and $q^R(x,t)$ are given by a series of eigenfunctions with eigenvalues that satisfy a transcendental equation.  The classical series solution may be obtained from the solution in Proposition~\ref{prop:2f} by deforming the contours along $\partial D_0^-$ and $\partial D_0^+$ to the real line, including small semi-circles around each root of either $\Delta_L(k)$ or $\Delta_R(k)$, depending on whether $q^L(x,t)$ or $q^R(x,t)$ is being calculated. Indeed, careful calculation of all different contributions, following the examples in~\cite{DeconinckTrogdonVasan, FokasBook, TrogdonDecon}, is allowed since all integrands decay in the wedges between these contours and the real line, and the zeros of $\Delta_L(k)$ and $\Delta_R(k)$ occur only on the real line, as stated above.  It is not necessarily beneficial to leave the form of the solution in Proposition~\ref{prop:2f} for the series representation, as the latter depends on the roots of $\Delta_L(k)$ and $\Delta_R(k)$, which are not known explicitly. In contrast, the representation of Proposition~\ref{prop:2f} depends on known quantities only and may be readily computed, using one's favorite parameterization of the contours $\partial D_0^-$ and $\partial D_0^+$.

\item In the case of the heat equation on the finite interval with an interface there are also an infinite number of poles on the real-$k$ axis.  The major difference here is that the boundary of $D$ coincides with the real-$k$ axis whereas in the heat equation the only intersection between the real axis and $D$ is at $k=0$.
%
\edits{\item As stated earlier, this method applies to general boundary conditions although we chose to present the details only for the Dirichlet case.  When genuine Robin boundary conditions are used (the case in which all of the coefficients $\alpha_1$, $\alpha_2$, $\alpha_3$, and $\alpha_4$ are nonzero) the analogue denominator to~\eqref{deltaL} may have zeros on the interior of $D_0^+$ and $D_0^-$ depending on the relative signs of the coefficients.  Thus, special care is needed to eliminate unknown boundary values in these cases.  This can be worked out in a straightforward way, as is done for problems without interfaces~\cite{DeconinckTrogdonVasan}.
}

\item Similar to Section~\ref{sec:ii}, long time asymptotics are easily computed using the method of stationary phase~\cite{BenderOrszag}.  The asymptotic behavior is centered around zero for $x/t$ constant and the shape of the envelope is determined by the integrands of the solution given in Proposition~\ref{prop:2f} as in~\eqref{2i_Lsoln_LOB} and~\eqref{2i_Rsoln_LOB}.

\end{itemize}

\section*{Acknowledgements}

This work was generously supported by the National Science Foundation under grant NSF-DMS-1008001 (B.D.).  N.E.S. also acknowledges support from the National Science Foundation under grant number NSF-DGE-0718124.  Any opinions, findings, and conclusions or recommendations expressed in this material are those of the authors and do not necessarily reflect the views of the funding sources.

\end{document}